\documentstyle[12pt,cite,epsf,axodraw]{article}
\title{Two loop radion correction to $K_L$ - $K_S$ mass difference 
in the stabilised Randall-Sundrum brane world scenario}

\author{Abhinav Gupta\thanks{E--mail : abh@ducos.ernet.in} and 
Namit Mahajan\thanks{E--mail : nm@ducos.ernet.in, 
nmahajan@physics.du.ac.in}\\
	{\em Department of Physics and Astrophysics,} \\
	 {\em University of Delhi, Delhi-110 007, India.}}

\begin{document}
\maketitle
\large
\begin{abstract}
In the stabilised Randall-Sundrum brane world scenario, the radion can
 have phenomenologically testable effects, which can be measured
 against precisely measured electroweak physics data. We investigate
 the effect of two loop radion corrections to  $K_L$ - $K_S$ mass
 difference to set a bound on the radion mass and vacuum expectation
 value. It is found that the leading two loop corrections 
are of the order $\left[Log\left(\frac{\Lambda^2}{m_\phi^2}\right)
\right]^2$ where $\Lambda$ is the cut-off scale ${\cal O}$($\sim$TeV) 
and $m_\phi$ is the radion mass.   
\end{abstract} 
\vskip 1cm
\begin{section}*{Introduction}
\indent Recently, the idea that the fundamental scale of quantum 
gravity could be dramatically low (${\cal O}\sim$ TeV) as opposed 
to the Planck scale, if the Standard Model (SM) fields lie on a
3-brane while gravity propagates in the bulk, has gained a lot of 
attention \cite{nima}-\cite{randall}. This is yet another attempt
to solve the hierarchy problem that plagues the SM. 
Arkani-Hamed, Dimopoulos and Dvali \cite{nima} proposed the existence
of $n$ additional large compact spatial dimensions in which gravity 
propagates while the SM fields are confined to 3-brane. The usual four
 dimensional Planck scale ($M_{pl}$) is related to the the effective 
Planck scale in the bulk (the String scale,$M_s \sim {\cal O}$(TeV)) and 
the volume of the compactified space ($V_n$) as 
\begin{equation}
M_{pl}^2 \sim V_n~M_s^{n+2}
\end{equation}
The bulk space-time is the direct product of four dimensional Minkowski   
space and the compact space, thus resulting in the relation as in 
equation(1). Thus, a large value of $V_n$ may considerably lower the 
effective scale of quantum gravity, making it possible
to test the predictions of such a theory. The imposed requirement of 
the radius of extra dimensions being large makes such a scenario less
attractive.
\indent Randall and Sundrum \cite{randall} suggested a different picture
in which the extra dimensions are small but the background metric is 
not flat. 
In the Randall-
Sundrum scenario, the SM fields live on one of the two 3-branes in
a five dimensional non-factorizable bulk space-time (a slice of 
$AdS_5$ space). The metric is
\begin{equation}
ds^2 = e^{-2kr_c|y|}\eta_{\mu\nu}dx^{\mu}dx^{\nu} - r_c^2dy^2
\end{equation}
where $x^{\mu}$ are the 4-dimensional coordinates while 
$y \in [-\pi,\pi]$ parameterizes a $S^1/Z_2$ orbifold. 
$r_c$ is the radius and $k^{-1}$
is the curvature radius of the $AdS_5$ space. The two branes lie at
$y = 0$ and $y = \pi$ with the later brane identified as the 'visible' 
brane.    

The geometry of the embedding space-time
 introduces an exponential warp factor, resulting in a hierarchy
 between mass scales on the two branes, lowering the natural scale of
 quantum gravity on the so called `visible' brane, which happens to 
be our universe.  
The quantum fluctuations of the inter-brane separation (the radius, $r_c$,
of the extra dimension)
 manifest themselves in the form of a scalar particle, the radion,
 whose mass can be quite low, making it possible for it to have a
 phenomenologically viable impact, even at low energies. 
The issue of stabilizing this inter-brane distance or modulus, is 
an important one.
Goldberger 
and Wise suggested a simple mechanism of stabilising this radius
wherein, an additional bulk scalar field coupling to both the branes, 
is employed \cite{wise}, providing a nontrivial potential for the radion,
consequently stabilizing it. In such a stabilizing scenario, the radion
field has a mass which is smaller than the lowest lying Kaluza-Klein mode 
of the graviton. 
\\
\indent The induced metric on the visible brane is
\begin{equation}
g_{\mu\nu}^{ind}(x) = e^{-\frac{\phi(x)}{\langle\phi\rangle}}
~g_{\mu\nu}(x)
\end{equation}
where $\phi$ is the radion field, $\langle\phi\rangle$ is it's 
vacuum expectation
 value (VEV) and $g_{\mu\nu}(x) = \eta_{\mu\nu} + h_{\mu\nu}(x)$ gives 
a tower of massive Kaluza-Klein gravitons as fluctuations about the
 flat Minkowski metric $\eta_{\mu\nu} = diag(1,-1,-1,-1)$, with the 
zero mass mode coupling to the SM fields with the usual gravitational
 strength and the massive modes coupling with a strength comparable to
 the weak scale. \\
\indent  As far as only radion corrections are concerned, the induced metric
 takes the form 
\begin{equation}
g_{\mu\nu}^{ind}(x) = e^{-\frac{\phi(x)}{\langle\phi\rangle}}~
\eta_{\mu\nu}
\end{equation}
The matter action on the visible brane takes the form
\begin{equation}
S_{matter} = \int d^4x~\sqrt{-g_{ind}}~\mathcal{L_{SM}}
\end{equation}
with $\mathcal{L_{SM}}$ being the SM Lagrangian in this warped
background. At the moment, one can lay bounds on radion 
parameters from precisely measured quantities like the 
electroweak oblique parameters \cite{csaba} and unitarity constraints
arising from gauge boson scattering \cite{uni}.
In this paper, we intend to constrain the radion parameters
using $K_L$-$K_S$ mass difference data.
\end{section}

\begin{section}*{Radion Coupling to the SM fields}
We derive the relevant Feynman rules for radion interacting 
with the SM particles from the first principles.  We first
consider for simplicity a simple gauge theory with a fermion and a
 massive gauge boson (both having acquired mass through Higgs
 mechanism) in the above warped background. Using the vierbein 
formalism for the fermions, the Lagrangian for the interaction of the
 radion with the fermion and the gauge boson is
\begin{eqnarray}
\mathcal{L} &=& e^{-3\phi \over {2\langle\phi\rangle}}~(\iota~\bar{\psi}
\gamma^\mu\partial_\mu\psi + g~\bar{\psi}\gamma^\mu A_\mu\psi) 
- e^{-2\phi \over {\langle\phi\rangle}}~(m\bar{\psi}\psi)\\ \nonumber 
&-& \frac{1}{4}~F^{\mu\nu}F_{\mu\nu} + 
e^{-\phi \over {\langle\phi\rangle}}
~ (\frac{1}{2}~M^2A^\mu A_\mu)
\end{eqnarray}
where g is the gauge coupling. In deriving the Feynman rules, one has
 to remember that it is the interaction Hamiltonian and not the
 Lagrangian that enters the S-matrix. The presence of a warp factor in
 the kinetic part of the fermion changes the canonical momentum from
 the usual flat background form by a multiplicative factor 
$e^{-3\phi \over {2\langle\phi\rangle}}$. This results in the interaction 
Hamiltonian being different from that obtained by naively replacing 
${\mathcal{H}}_{int}$ by $-{\mathcal{L}}_{int}$ for the fermion 
(see Appendix for details).  \\
\indent If instead, the fermion field is rescaled as
\begin{equation}
\psi \longrightarrow e^{-3\phi \over {4\langle\phi\rangle}}~\psi
\end{equation}
the Lagrangian, in terms of this new field, reads
\begin{eqnarray}
\mathcal{L} &=& \iota~\bar{\psi}
\gamma^\mu\partial_\mu\psi + g~\bar{\psi}\gamma^\mu A_\mu\psi 
- e^{-\phi \over {2\langle\phi\rangle}}~(m\bar{\psi}\psi)\\ \nonumber 
&-& \frac{1}{4}~F^{\mu\nu}F_{\mu\nu} + 
e^{-\phi \over {\langle\phi\rangle}}
~ (\frac{1}{2}~M^2A^\mu A_\mu)
\end{eqnarray}
Now that there are no derivative terms involved and the kinetic energy
term appears in the canonical form, 
\begin{eqnarray}
{\mathcal{H}}_{int} &=& -{\mathcal{L}}_{int} \\ \nonumber
&=& e^{-\phi \over {2\langle\phi\rangle}}~
(m\bar{\psi}\psi) 
- e^{-\phi \over {<\phi>}}~ (\frac{1}{2}~M^2A^\mu A_\mu)
- g~\bar{\psi}\gamma^\mu A_\mu\psi
\end{eqnarray}
Clearly, there is no fermion-fermion-gauge boson-radion vertex which
appears in other similar calculations \cite{kim}.
The radion thus couples only 
to the mass terms and not to the terms involving momenta.
The generalization to SM is straight forward and gives the following
Feynman rules  
\vskip 3cm
\begin{figure}[htb]
\vspace*{-11ex}
\hspace*{2em}
\begin{tabbing}
\begin{picture}(155,120)(-5.0,-20)
\ArrowLine(0,0)(45,45)
	\Text(20,30)[r]{$\psi_i$}
\ArrowLine(45,45)(0,90)
	\Text(20,60)[r]{$\bar{\psi_j}$}
\DashLine(45,45)(90,45){3}
	\Text(90,55)[r]{$\phi$}
\Text(165,45)[r]{$\iota~\frac{m}{2<\phi>}~\delta_{ij}$}
\end{picture}
\hskip 1.5cm
\begin{picture}(155,120)(-5.0,-20)
\ArrowLine(0,0)(45,45)
	\Text(20,30)[r]{$\psi_i$}
\ArrowLine(45,45)(0,90)
	\Text(20,60)[r]{$\bar{\psi_j}$}
\DashLine(45,45)(90,90){3}
	\Text(100,90)[r]{$\phi$}
\DashLine(45,45)(90,0){3}
	\Text(100,0)[r]{$\phi$}
\Text(165,45)[r]{$-\iota~\frac{m}{4<\phi>^2}~\delta_{ij}$}
\end{picture}
\end{tabbing}
\end{figure}
\vskip 2.5cm
 \begin{figure}[htb]
\vspace*{-10ex}
\hspace*{2em}
\begin{tabbing}
\begin{picture}(155,120)(-5.0,-20)
\Photon(0,0)(45,45){5}{5}
	\Text(10,25)[r]{$W_{\mu}$}
\Photon(45,45)(0,90){5}{5}
	\Text(10,65)[r]{$W_{\nu}$}
\DashLine(45,45)(90,45){3}
	\Text(90,55)[r]{$\phi$}
\Text(165,45)[r]{$-\iota~\frac{M_W^2}{<\phi>}~\eta_{\mu\nu}$}
\end{picture}
\hskip 1.5cm
\begin{picture}(155,120)(-5.0,-20)
\Photon(0,0)(45,45){5}{5}
	\Text(10,25)[r]{$W_{\mu}$}
\Photon(45,45)(0,90){5}{5}
	\Text(10,65)[r]{$W_{\nu}$}
\DashLine(45,45)(90,90){3}
	\Text(100,90)[r]{$\phi$}
\DashLine(45,45)(90,0){3}
	\Text(100,0)[r]{$\phi$}
\Text(165,45)[r]{$\iota~\frac{M_W^2}{<\phi>^2}~\eta_{\mu\nu}$}
\end{picture}
\end{tabbing}
\end{figure}
\vskip 1cm
where i, j are the fermion flavour indices and m the fermion mass.\\
\indent The radion, therefore, seems to couple just like the SM Higgs
 up to first order in the relevant coupling. However, the underlying
 gauge symmetry of SM regulates the ultraviolet behaviour of radiative 
corrections due to the Higgs. This feature is absent in the radion 
interactions, giving substantial contributions to the otherwise 
suppressed radiative processes. Moreover, the presence of new vertices
(as opposed to the SM Higgs) results in extra diagrams relevant at 
the desired order in coupling.
\end{section}
\begin{section}*{Corrections to $\Delta m_K$}
The SM more or less predicts the $K^0$-$\bar{K^0}$ mixing to the 
observed level. However, the $\Delta m_K$ prediction is not so precise.
The QCD improved short-distance contributions arising from the familiar
box diagrams for the transition $s\bar{d} \longrightarrow \bar{s}d$
lead to the following effective Hamiltonian \cite{dono}
\begin{eqnarray}
{\mathcal H}_W^{Box} &=& \frac{G_F^2}{4\pi^2}~[\xi_c^2
S_0(x_c)\eta_1 + \xi_t^2S_0(x_t)\eta_2 \\ \nonumber
&+& 2\xi_c\xi_tS_0(x_c,x_t)\eta_3]~(\bar{d}_L
\gamma^\alpha s_L)~(\bar{d}_L\gamma_\alpha s_L) + h.c.  
\end{eqnarray}
where $\xi_i=V_{id}^{\ast}V_{is}$ is the CKM factor with the index 
$i$ running over the quark flavours u, c and t  and  $x_i = 
\frac{m_i^2}{M_W^2}$
\begin{equation}
S_0(x) = \left[\frac{1}{4} + \frac{9}{4(1 - x)} - \frac{3}{2(1 - x)^2}
\right] - \frac{3x^2}{2(1 - x)^3}~lnx
\end{equation}
\begin{eqnarray}
S_0(x,y) &=& y\Bigg[-\frac{1}{y - x}\left(\frac{1}{4} +
 \frac{3}{2(1 - x)} - \frac{3}{2(1 - x)^2}\right)~lnx \\ \nonumber
&+& (y \leftrightarrow x) - \frac{3}{4(1 - x)(1 - y)}\Bigg]
\end{eqnarray}
with $\eta_i$ representing the QCD effects. \\
\indent We now calculate the radion correction to $\Delta m_K$. 
The relevant box diagrams for the process $s\bar{d} \longrightarrow 
\bar{s}d$ are 
\vskip 2.0cm
 \begin{figure}[htb]
\vspace*{-11ex}
\hspace*{2em}
\begin{tabbing}
\begin{picture}(155,120)(-5.0,-20)
\Line(0,0)(150,0)
	\Text(5,10)[c]{d}
	\Text(145,10)[c]{s}
	\Text(65,10)[c]{j}
	\Text(65,80)[c]{i}
\Line(0,90)(150,90)
	\Text(5,100)[c]{s}
	\Text(145,100)[c]{d}
\DashLine(30,45)(120,45){5}
	\Text(75,55)[c]{$\phi$}
\Photon(30,90)(30,0){5}{8}
	\Text(10,45)[l]{W}
\Photon(120,90)(120,0){5}{8}
	\Text(140,45)[r]{W}
	\Text(75,-10)[c]{(a)}
\end{picture}
\hskip 1.5cm
\begin{picture}(155,120)(-5.0,-20)
\Line(0,0)(150,0)
	\Text(5,10)[c]{d}
	\Text(145,10)[c]{s}
	\Text(65,10)[c]{j}
	\Text(65,80)[c]{i}
\Line(0,90)(150,90)
	\Text(5,100)[c]{s}
	\Text(145,100)[c]{d}
\DashLine(75,90)(75,0){5}
	\Text(85,45)[c]{$\phi$}
\Photon(30,90)(30,0){5}{8}
	\Text(10,45)[l]{W}
\Photon(120,90)(120,0){5}{8}
	\Text(140,45)[r]{W}
	
	\Text(75,-10)[c]{(b)}
\end{picture}
\end{tabbing}
\end{figure}
\vskip 3.0cm
 \begin{figure}[htb]
\vspace*{-11ex}
\hspace*{2em}
\begin{tabbing}
\begin{picture}(155,120)(-5.0,-20)
\Line(0,0)(150,0)
	\Text(5,10)[c]{d}
	\Text(145,10)[c]{s}
	\Text(65,10)[c]{j}
	\Text(65,80)[c]{i}
\Line(0,90)(150,90)
	\Text(5,100)[c]{s}
	\Text(145,100)[c]{d}
\DashCArc(150,90)(65,180,245){10}
	\Text(75,55)[c]{$\phi$}
\Photon(30,90)(30,0){5}{8}
	\Text(10,45)[l]{W}
\Photon(120,90)(120,0){5}{8}
	\Text(140,45)[r]{W}
	\Text(75,-10)[c]{(c)}
\end{picture}
\hskip 1.5cm
\begin{picture}(155,120)(-5.0,-20)
\Line(0,0)(150,0)
	\Text(5,10)[c]{d}
	\Text(145,10)[c]{s}
	\Text(65,10)[c]{j}
	\Text(65,80)[c]{i}
\Line(0,90)(150,90)
	\Text(5,100)[c]{s}
	\Text(145,100)[c]{d}
\DashCArc(75,90)(30,0,180){5}
	\Text(100,120)[c]{$\phi$}
\Photon(30,90)(30,0){5}{8}
	\Text(10,45)[l]{W}
\Photon(120,90)(120,0){5}{8}
	\Text(140,45)[r]{W}
	
	\Text(75,-10)[c]{(d)}
\end{picture}
\end{tabbing}
\end{figure}
\vskip 2.0cm
 \begin{figure}[htb]
\vspace*{-11ex}
\hspace*{2em}
\begin{tabbing}
\begin{picture}(155,120)(-5.0,-20)
\Line(0,0)(150,0)
	\Text(5,10)[c]{d}
	\Text(145,10)[c]{s}
	\Text(65,10)[c]{j}
	\Text(65,80)[c]{i}
\Line(0,90)(150,90)
	\Text(5,100)[c]{s}
	\Text(145,100)[c]{d}
\DashCArc(75,115)(25,0,360){10}
	\Text(105,115)[c]{$\phi$}
\Photon(30,90)(30,0){5}{8}
	\Text(10,45)[l]{W}
\Photon(120,90)(120,0){5}{8}
	\Text(140,45)[r]{W}
	\Text(75,-10)[c]{(e)}
\end{picture}
\hskip 1.5cm
\begin{picture}(155,120)(-5.0,-20)
\Line(0,0)(150,0)
	\Text(5,10)[c]{d}
	\Text(145,10)[c]{s}
	\Text(65,10)[c]{j}
	\Text(65,80)[c]{i}
\Line(0,90)(150,90)
	\Text(5,100)[c]{s}
	\Text(145,100)[c]{d}
\DashCArc(120,45)(30,270,90){5}
	\Text(145,22)[c]{$\phi$}
\Photon(30,90)(30,0){5}{8}
	\Text(10,45)[l]{W}
\Photon(120,90)(120,0){5}{8}
	\Text(140,45)[r]{W}
	
	\Text(75,-10)[c]{(f)}
\end{picture}
\end{tabbing}
\end{figure}
\vskip 2.0cm
 \begin{figure}[htb]
\vspace*{-11ex}
\hspace*{2em}
\begin{tabbing}
\begin{picture}(155,120)(-5.0,-20)
\Line(0,0)(150,0)
	\Text(5,10)[c]{d}
	\Text(145,10)[c]{s}
	\Text(65,10)[c]{j}
	\Text(65,80)[c]{i}
\Line(0,90)(150,90)
	\Text(5,100)[c]{s}
	\Text(145,100)[c]{d}
\DashCArc(150,45)(25,0,360){10}
	\Text(180,40)[c]{$\phi$}
\Photon(30,90)(30,0){5}{8}
	\Text(10,45)[l]{W}
\Photon(120,90)(120,0){5}{8}
	\Text(140,45)[r]{W}
	\Text(75,-10)[c]{(g)}
\end{picture}
\end{tabbing}
\end{figure}
and their left-right or up-down reflections.\\

\indent The calculations have been performed in the unitary gauge using 
dimensional regularization. The GIM mechanism ensures that the 
integrals are logarithmically divergent, thus justifying the use of
 dimensional regularization. The choice of the Unitary gauge avoids 
the inclusion of ghosts and additional corrections due to the 
interaction of the radion with the gauge fixing term, a point missed 
in\cite{kim}. The two-loop integrals (at zero external momenta) have 
been evaluated in $d = 4-\epsilon$ dimensions using the techniques 
similar to those given in \cite{veltman}.
 It is found that the dominant contribution (of the form 
$\frac{1}{\epsilon^2}$) comes from the diagrams (b), (c) and (d).
 The total leading 
contribution to the amplitude is 
\begin{eqnarray}
\iota~T_{radion}(s\bar{d} \rightarrow \bar{s}d) &=&\iota~ 
\frac{G_F}{\sqrt{2}}~\frac{\alpha}{\pi\sin^2\theta_w} 
~\left(\frac{1}{16\pi^2}\right)~\frac{3M_W^2}{8\langle\phi\rangle^2}~
\frac{1}{\epsilon^2}\\ \nonumber
&(&\bar{d}_L\gamma^\alpha s_L)~(\bar{d}_L\gamma_\alpha s_L)
~\sum_{i,j}~x_ix_j~\xi_i\xi_j
\end{eqnarray}
where $G_F$ is the Fermi constant, $\alpha$ the fine structure 
constant, $\theta_w$ the Weinberg angle. In the above 
expression, it is understood that $\frac{1}{\epsilon}$ is to be
 replaced by $Log\left(\frac{\Lambda^2}{m_\phi^2}\right)$, where
 $\Lambda$ is the cut-off scale of the theory. This results in the
 correction to $\Delta m_K$ being given by
\begin{eqnarray}
\Delta m_K^{radion} &=& -\frac{G_F}{\sqrt{2}}~\frac{\alpha}{6\pi}~
\frac{f_K^2m_K}{\sin^2\theta_w} ~\left(\frac{1}{16\pi^2}\right)~
\frac{3M_W^2}{8\langle\phi\rangle^2}~\left[Log\left(\frac{\Lambda^2}
{m_\phi^2}\right)\right]^2\\ \nonumber 
&~&Re(\sum_{i,j}~x_ix_j~\xi_i\xi_j)
\end{eqnarray}
with $f_K$ and $m_K$ being the Kaon decay constant and mass respectively. \\ 
\indent As mentioned before, the observed size of 
$\Delta m_K$ ($(3.489\pm 0.009)\times 10^{-12}$ MeV) \cite{pdg}
 is roughly compatible
 with that expected in SM.
Therefore, it would be natural to constrain the parameters of the 
theory (radion mass and VEV) by imposing the condition 
\begin{equation}
\Delta m_K^{radion} \leq 0.009~\times 10^{-12}~ MeV  
\end{equation}
This equation imposes an inequality on the space spanned by $m_\phi$
and $\langle\phi\rangle$. constraining the {\it allowed region} 
(the region in the $m_\phi$-$\langle\phi\rangle$ plane above the
 curve in Fig.1).
\end{section}

\begin{section}*{Appendix}
Here we outline the steps that lead to the expression for 
interaction Hamiltonian as in equation(7) starting from 
the Lagrangian given in equation(6).
Denoting the momentum conjugate to the fermion field $\psi$ by 
$\Pi_{\psi}~ = ~\frac{\partial{\mathcal{L}}}{\partial(\partial_0\psi)}$
we have from equation(6)
\begin{equation}
\Pi_{\psi} = e^{-3\phi \over {2\langle\phi\rangle}}~\iota~\psi^{\dag}
\end{equation}
which is different from the corresponding flat space-time expression
by the multiplicative warp factor while the momentum conjugate to the gauge
field, $\Pi_A$, is same as in the flat space-time case.
The Hamiltonian is
\begin{equation}
{\mathcal{H}} = \Pi_{\psi}~\partial_0\psi + \Pi_A~\partial_0A - \mathcal{L}
\end{equation}
with $\mathcal{L}$ as given in equation(6) and the Lorentz indices 
for the gauge field and its momenta suppressed.
The interaction Hamiltonian is just the terms left after subtracting the 
free Hamiltonian from the above Hamiltonian.
Making a canonical transformation 
to the interaction representation, the fields and momenta are changed into
$\psi_{in}$, $A_{in}$, ${\Pi_A}_{in}$
 and ${\Pi_{\psi}}_{in}~=~\iota\psi_{in}^{\dag}$. 
Therefore, in
the interaction representation (dropping the subscript $in$),
\begin{equation}
{\mathcal{H}}_{int} = e^{-\phi \over {2\langle\phi\rangle}}~
(m\bar{\psi}\psi) 
- e^{-\phi \over {<\phi>}}~ (\frac{1}{2}~M^2A^\mu A_\mu) - 
g~\bar{\psi}\gamma^\mu A_\mu\psi
\end{equation}
which is exactly the same as obtained by rescaling the fermion field.
\end{section}

\begin{section}*{Acknowledgements}
The authors thank S.~Rai Choudhury for discussions and suggestions.
 A.~G thanks CSIR, India while N.~M. thanks the 
University Grants Commission, India, for fellowship.
\end{section}

\pagebreak
\begin{figure}[ht]
\vskip 15truecm
\includegraphics{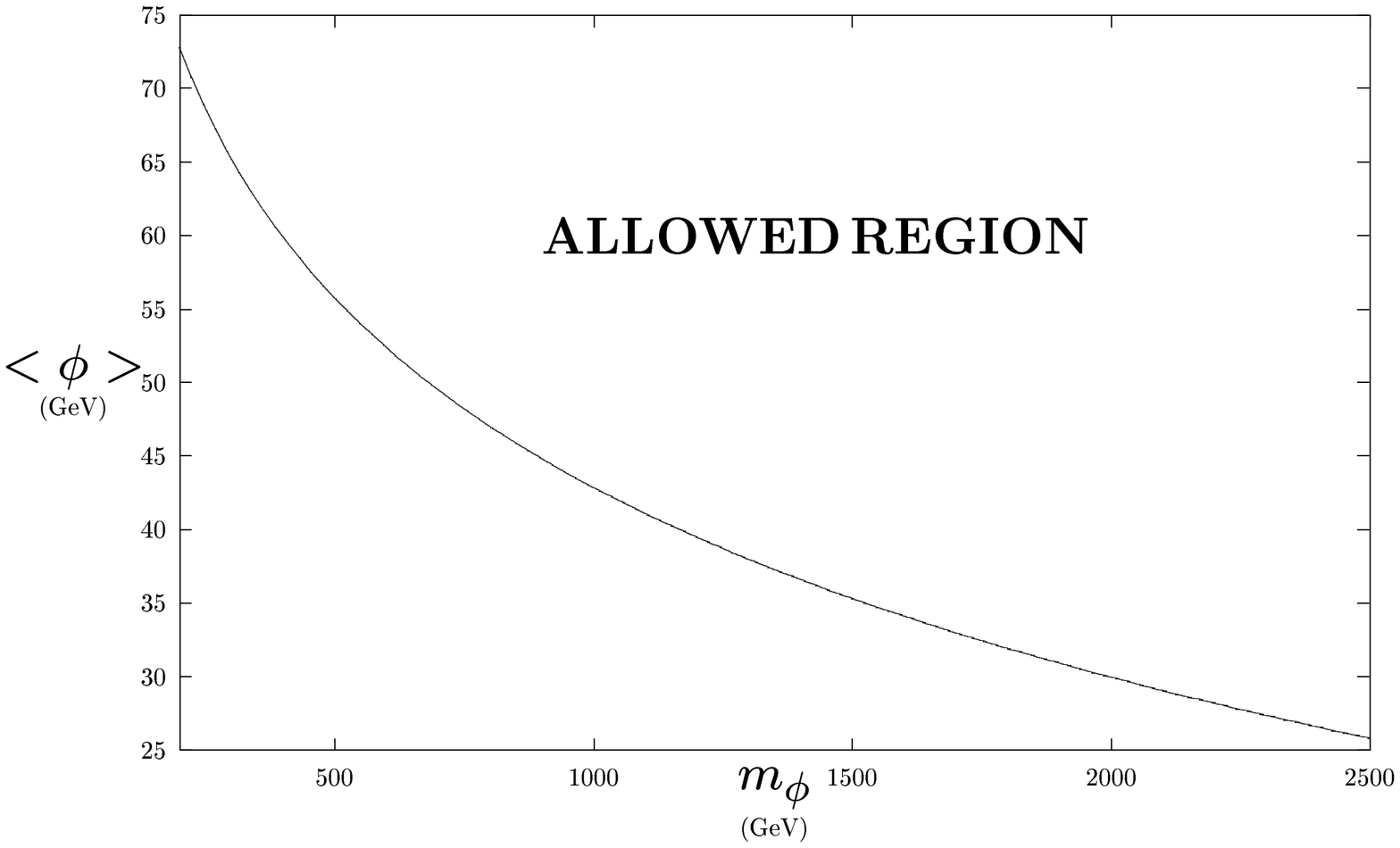}
\begin{center}
Fig.1: The allowed region for the vacuum expectation value and mass of the 
radion for cut-off $\Lambda~=~10$ TeV.
\end{center}

\end{figure}
\end{document}